\documentclass[twocolumn,prl,aps,superbib,tightenlines,floatfix]{revtex4}
\usepackage{amsfonts}
\usepackage{amsmath}
\usepackage{amssymb}
\usepackage{graphicx}

\newcommand{\dIs}{\mathrm{d}^2I/\mathrm{d}V^2}

\begin{document}

\title{Inelastic effects on the transport properties of alkanethiols}
\author{$^{1}$Yu-Chang Chen, $^{2}$Michael Zwolak, and $^{1}$Massimiliano Di
Ventra}
\affiliation{$^{1}$Department of Physics, University of California, San Diego, La Jolla,
CA 92093-0319}
\affiliation{$^{2}$Physics Department, California Institute of Technology, Pasadena,
CA 91125}

\begin{abstract}
Using first-principles approaches we investigate 
local heating and the inelastic contribution to the current for various alkanethiols 
sandwiched between metal electrodes. In the absence of good heat dissipation into the bulk
electrodes, we find that the local temperature of the alkanethiols is relatively
insensitive to their length. This is due to the rates of heating and cooling processes scaling 
similarly with length. On the other hand, when considering heat dissipation into the bulk
electrodes, the local temperature of alkanethiols decreases as their
length increases. We also find that the inelastic scattering profile displays 
an odd-even effect with length which compares well with experimental 
results. This effect is due to the alternating direction
of the CH$_3$ group motion with respect to current flow with increasing C atoms in the chain, and 
is very sensitive to the structure of the carbon-sulfur-gold bond. 
Inelastic scattering profiles can therefore help illuminate the bonding configuration 
of molecules to metallic surfaces. 
\end{abstract}

\maketitle

There is an ever-increasing interest in charge transport in organic molecules due to 
their potential application in electronic devices.~\cite{Xu,Reed,DiVentra2000,Nitzan,Nazin,Yu,Reichert} 
Recently, several laboratories have reported consistent data in the resistance of alkyl 
chains.~\cite{Zhao,Wold,Cui2,Beebe,Holmlin} Some theoretical calculations support 
these results,~\cite{Guo} at least in the shape of the current-voltage (I-V) characteristics.~\cite{dicarlo,Cui} 
This seems to suggest that a reproducible contact can be created between the alkanethiols and the 
electrodes.~\cite{Cui,Kushmerick,Wang1,Guo,Zhao,Wang2,Kluth}  
However, several current-induced mechanical effects such as forces on 
ions~\cite{DiVentra04} and local heating~\cite{Todorov2,DiVentra03,smit} 
can generate substantial structural instabilities which can lead to atomic geometries quite different than those assumed theoretically. 

In this paper, we focus on one current-induced mechanical effect, namely, inelastic scattering in alkanethiol molecular junctions.
Our intent is to explore (i) the dependence of local heating on the length of the alkyl chains and gain insight into their stability 
under current flow, and (ii) determine the inelastic contribution to the current. We find that the local temperature
of alkyl chains is smaller the longer the chain provided that there is good thermal dissipation into the bulk electrodes.
This is due to the insulating character of the alkanethiols
and is in contrast to results for metallic quantum-point contacts.~\cite{DiVentra04,Todorov2}  
We also find that the inelastic scattering profile displays an odd-even effect which  
is very sensitive to the structure of the carbon-sulfur-gold bond. This odd-even effect is in agreement with 
high resolution electron energy loss spectroscopy (HREELS) experiments on the same systems.~\cite{Kato} This demonstrates that inelastic spectroscopy can 
be quite effective in determining the atomic-scale geometry of molecular junctions.~\cite{cheninelastic} 

\begin{figure}
\includegraphics[width=.48\textwidth]{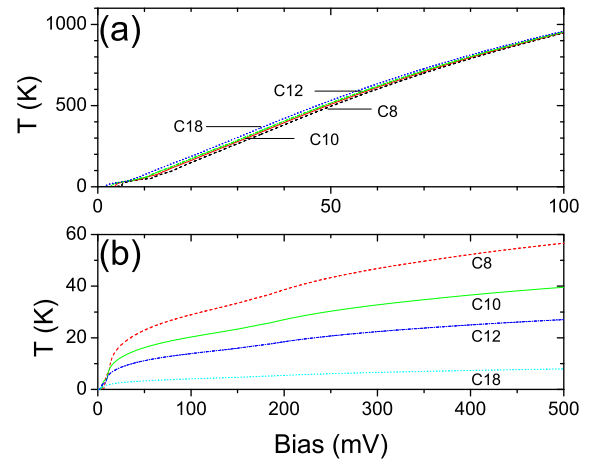}
\caption{Local temperature as a function of bias for various lengths of
alkanethiols. (a) No heat dissipation into the bulk electrodes is taken into
account. (b) Heat dissipation into the bulk electrodes is taken into
account.}
\label{fig1}
\end{figure}

We investigate these inelastic effects in the framework of static density-functional theory (DFT) in the local density
approximation.~\cite{diventra} The calculations proceed as follows: first, the stationary scattering wavefunctions of ethanethiolate
sandwiched between two bulk gold electrodes (represented with ideal metals, jellium model,
r$_{\text{s}}\approx $ 3) are
calculated by solving the Lippmann-Schwinger equation self-consistently.~\cite{diventra} 
Due to their insulating character, the current in these systems decreases exponentially with length $d$ as 
$I=I_{0}\exp (-\beta d)$.~\cite{Wang1,Guo,Beebe,prec} By exploiting the periodicity in (CH$_2$)$_2$ 
groups of the alkyl chains, we can then calculate the wavefunctions of the different molecules by a 
simple scaling argument. The vibrational mode energies and the transformation matrix 
which contains the character of the modes (longitudinal versus transverse with 
respect to current flow)~\cite{DiVentra03} are evaluated using 
total-energy calculations.~\cite{vibmodes} Throughout the paper, the angle that the 
S-C bond makes with the surface normal has been fixed at 43 degrees, which was obtained by relaxing the structures at 
zero bias. This angle is in reasonable agreement with the one found in previous theoretical 
work.~\cite{Morikawa} Below we will show the sensitivity of the inelastic current to the variation of this 
angle. The electron-phonon coupling constant for each atom and mode of the chain is finally  
calculated as reported in Ref.~\onlinecite{DiVentra03}. With these quantities we can then evaluate both the local temperature of the 
junction and the inelastic I-V curve. 

\begin{figure}
\includegraphics[width=.48\textwidth]{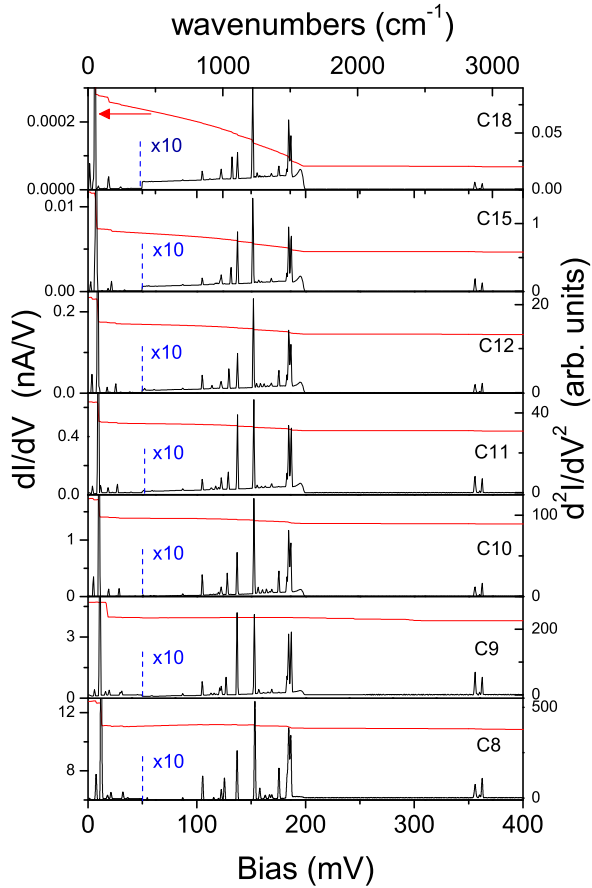}
\caption{Differential conductance (left axis) and absolute value of 
$\dIs$ (right axis) as a function of bias for various alkanethiols
with different numbers of carbon atoms. Due to the large relative magnitude of the inelastic features of low-energy modes 
with respect to the high-energy modes, the curve $\dIs$ has been multiplied by a factor of ten for biases larger than 50 mV.} 
\label{IETS}
\end{figure}

\textit{Local heating} - Let us first investigate the effect of heating due to electron-phonon 
interaction. This interaction allows for the exchange of energy between electrons and the lattice via
absorption and emission of vibrational modes. Details of the
theory of local heating in nanoscale structures can be found in
Refs.~\onlinecite{DiVentra03,Todorov1}. Here we just mention
that there are two major inelastic channels that lead to a given local
temperature in a nanojunction. One is due to inelastic processes
that occur in the atomic region of the junction, i.e., involving the atoms of the alkanethiols and 
few atomic layers of the bulk electrodes. Electrons incident from the right or left electrode can absorb (cooling) or
emit (heating) energy because of electron-vibration scattering with the ions of 
the junction. The other channel is due to dissipation of energy into the
bulk electrodes via elastic phonon scattering. In quasi-ballistic systems, where the 
inelastic electron mean free path is large compared to the dimensions of the junction,
elastic phonon scattering is the most efficient way of dissipating heat into the bulk electrodes.~\cite{DiVentra03,Todorov1} 
Let us first focus on the inelastic scattering contribution
assuming heat dissipation into the electrodes is negligible. This can be the result
of, e.g., weak coupling of vibrational modes localized in the
junction with the continuum of modes of the bulk
electrodes.~\cite{DiVentra03,Todorov2} 

Denoting by $W_{\nu}^{L(R),1(2)}$ the power absorbed (emitted) by electrons incident
from the left (right) via a vibrational mode $\nu$, the total
thermal power generated in the junction can be written as the sum
over all vibrational modes of the above four scattering
processes~\cite{cheninelastic}:
\begin{equation}
P=\sum_{\nu \in vib.}\left( W_{\nu }^{R,2}+W_{\nu }^{L,2}-W_{\nu
}^{R,1}-W_{\nu }^{L,1}\right)  \label{power}
\end{equation}
This power can be expressed in terms of
the electron-vibration coupling in the presence of current.~\cite{cheninelastic}
When the heating processes ($W_{\nu }^{R,2}$\ and $W_{\nu }^{L,2}$) balance the cooling
processes ($W_{\nu }^{R,1}\ $and $W_{\nu }^{L,1}$), i.e., $P=0$, a
steady-state local temperature is established in the junction.
This temperature is plotted in Fig.~\ref{fig1}(a) for
various alkanethiols of different length assuming zero
background temperature. It is clear from the figure that the local temperature
without heat dissipation depends weakly on the length of the molecule. This can be understood quite 
easily since the rate of
energy transfer $W_{\nu }^{\alpha ,k}$ scales similarly with length for all modes
and processes (heating and cooling). In other words, even though the power {\it per mode} decreases
exponentially with length by a factor $\exp (-\beta d)$ due to the corresponding exponential decrease of the 
current, the {\it total} power (Eq.~\ref{power}) will be zero at approximately the same temperature, as the above  
exponential term factors out. 

We now allow for the energy stored locally in the junction to dissipate 
away into the electrodes via the coupling between the normal modes of the molecule and the phonons of the 
electrodes. As in our previous work,~\cite{DiVentra03} we estimate this thermal conductance assuming the junction forms a weak mechanical link with a
given stiffness $K$.~\cite{geller} The thermal current into the electrodes via elastic phonon scattering is thus  
\begin{equation}
I_{th}=\frac{4\pi K^{2}}{\hbar }\int d\varepsilon \varepsilon N_{L}\left(
\varepsilon \right) N_{R}\left( \varepsilon \right) \left[ n_{L}(\varepsilon
)-n_{R}(\varepsilon )\right] , 
\label{thermal}
\end{equation}
where $n_{L(R)}$ is the Bose-Einstein distribution function and 
$N_{L(R)}\left( \varepsilon \right)$ is the spectral density of phonon states at
the left (right) electrode surface. Since the stiffness $K$ is proportional to $d^{-1}$,~\cite{geller} 
the thermal current $I_{th}$ is proportional to $d^{-2}$. 
Therefore, while the local power in the junction (Eq.~\ref{power}) that needs to 
be dissipated decreases exponentially with length, the thermal power into the electrodes that allows for 
this dissipation (Eq.~\ref{thermal}) decreases only algebraically. 
This implies a lower temperature in the junction as a function of length [see Fig.~\ref{fig1}(b)]. 
We note that this trend for alkanethiols 
is opposite to that found in atomic wires~\cite{DiVentra04,Todorov2} and is due to their insulating 
character (conversely the current in atomic wires is relatively independent of length~\cite{DiVentra04}). 
This result combined with the fact that current-induced forces decrease with increasing length~\cite{yang2003,DiVentra04,prec2} 
suggests that longer alkanethiols are more stable against current flow. This is consistent with recent 
experimental results on similar systems.~\cite{Wold}

\begin{figure}
\includegraphics[width=.48\textwidth]{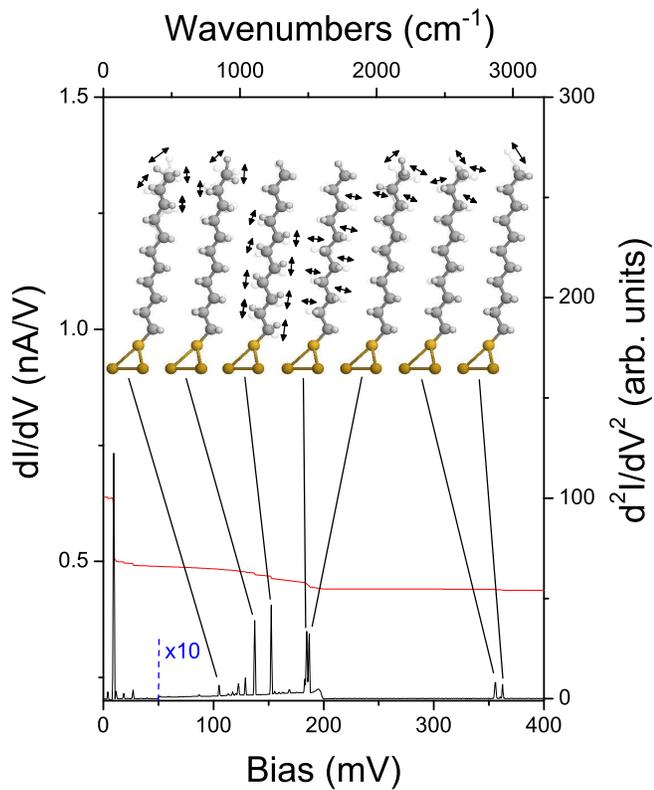}
\caption{Differential conductance (left axis) and absolute value of $\dIs$ (right axis) as a function of bias for undecanethiolate. 
The curve $\dIs$ has been multiplied by a factor of ten for biases larger than 50 mV.}
\label{C11}
\end{figure}

\begin{figure}
\includegraphics[width=.46\textwidth]{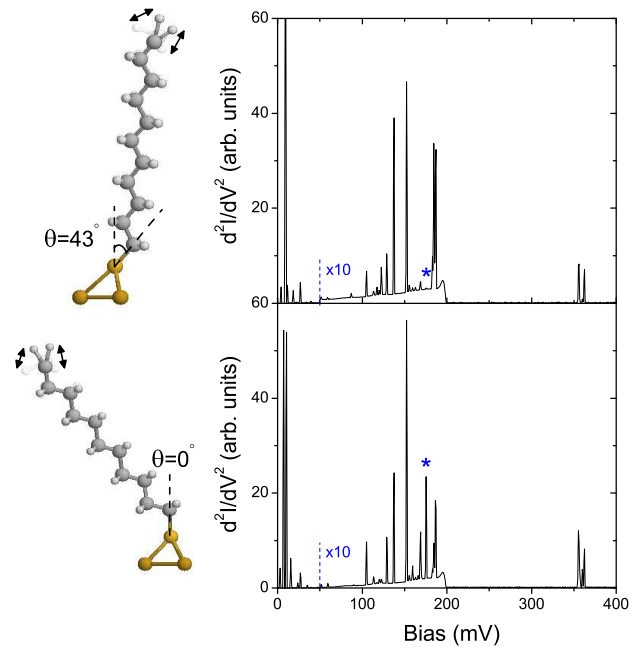}
\caption{Absolute value of $\dIs$ as a function of bias for two different surface-S-C angles of undecanethiolate. Top: 
the structure with an angle of 43 degrees. Bottom: the structure with an angle of zero degrees. The change in angle 
transforms the mode at 176 mV (indicated by an 
asterisk) from quasi-transverse (top-left schematic) to quasi-longitudinal (bottom-left schematic) so that it appears as a 
new feature in the inelastic conductance.}
\label{GeomEffect}
\end{figure}

\textit{Inelastic current - } The coupling between electrons and molecular
vibrations will cause small discontinuities in the conductance at biases corresponding to the energy of normal 
modes.~\cite{cheninelastic,Todorov2,troisi}
The position and magnitude of these discontinuities varies significantly from mode to mode and
can be measured in experiments such as inelastic electron tunneling
spectroscopy (IETS)~\cite{Kushmerick,Wang2} and HREELS.~\cite{Kato,Kluth} The theory of inelastic scattering as 
we use it in this paper is reported in Ref.~\onlinecite{cheninelastic}. We just recall here that the relative strength of 
inelastic features is mainly determined by the character of the modes: those that have large longitudinal character with respect 
to current flow contribute most to the features in the inelastic conductance.~\cite{cheninelastic,Todorov2,troisi} 

In Fig.~\ref{IETS} we plot the inelastic contribution to the conductance and 
its first derivative as a function of bias for various alkyl
chains. A broadening of 1 meV corresponding to elastic phonon scattering has been introduced to make the curve $\dIs$ 
finite. We show the modes that contribute the major peaks in Fig.~\ref{C11} for the case of undecanethiolate. 
The major vibrational modes in Fig.~\ref{IETS} are in good agreement
with the results of HREELS experiments on self-assembled monolayers.~\cite{Kato, Kluth} The differences between the reported inelastic curves 
and the theoretical ones can be partly ascribed to modification of vibrational modes due to 
molecule-molecule interaction, an effect which we do not consider in this work. Most of the peaks at small
bias ($V<50$ mV) are associated with stretching, bending, and twisting motions of the alkane backbone. The largest peak at low 
bias (also observed in IETS and HREELS experiments~\cite{Kushmerick,Kato}) corresponds to a rigid motion of the whole molecule with respect to the electrodes. 
We also observe a weak odd-even effect in the
spectrum of $\dIs$ at wavenumbers 850 cm$^{-1}$, 1107 cm$^{-1}$, and 1420 cm$^{-1}$. This effect is due to the alternating direction
of the CH$_3$ group motion with respect to current flow with increasing C atoms in the chain. It is most pronounced for 
the 1420 cm$^{-1}$ mode, which is in agreement with HREELS experiments.~\cite{Kato}

To conclude we discuss how inelastic spectroscopy can probe the atomic-scale geometry of the alkanethiol-gold
bond. In the above, we have considered 
a geometry where the S-C bond forms an angle of 43 degrees with the surface normal (see top-left schematic of Fig.~\ref{GeomEffect}). 
However, recent calculations have found different energetically stable structures depending on the details of the computational methods 
employed.~\cite{Morikawa,Gronbeck,Akinaga,Yourdshahyan,Vargas}  The actual geometry will be determined by experimental conditions, including the 
way the contacts are formed. In Fig.~\ref{GeomEffect} we show that a change of the above angle from 
43 degrees to, e.g., zero degrees can change substantially the character of the modes 
in the structure. In this case, the change in angle transforms the quasi-transverse  
character of the mode at 1420 cm$^{-1}$ to a quasi-longitudinal one (see Fig.~\ref{GeomEffect}). 
This mode is the one responsible for the most pronounced odd-even effect we discussed above. Therefore, 
with a surface-S-C bond angle of zero degrees, this effect would be reversed as a function of the number 
of C atoms in the chain. Inelastic spectroscopy, combined with this type of detailed 
calculations, can therefore help illuminate the bonding configuration of molecules in metal-molecule-metal structures. 

The authors would like to thank M. Chshiev for help in some of the calculations.  
We acknowledge support from the NSF Grant No. DMR-01-33075. MZ 
acknowledges support from an NSF Graduate Fellowship.


\begin{thebibliography}{99}

\bibitem{Xu} Xu, B.; Tao, N. J. {\it Science} {\bf 2003}, {\it 301}, 1221.

\bibitem{Reed} Reed, M. A.; Zhou, C.; Muller, C. J.; Burgin, T. P.; Tour, J. M.
{\it Science} {\bf 1997}, {\it 278}, 252.

\bibitem{DiVentra2000} Di Ventra, M.; Pantelides, S. T.; Lang, N. D. {\it Phys. Rev. Lett.} {\bf 2000}, {\it 84}, 979.

\bibitem{Nitzan} Nitzan, A.; Ratner, M. A. {\it Science} {\bf 2003}, {\it 300}, 1384.

\bibitem{Nazin} Nazin, G. V.; Qiu, X. H.; Ho, W. {\it Science} {\bf 2003}, {\it 302}, 77.

\bibitem{Yu} Yu, L. H.; Keane, Z. K.; Ciszek, J. W.; Cheng, L.; Stewart, M. P.; Tour, J. M.; Natelson, D. cond-mat/0408052.

\bibitem{Reichert} Reichert, J.; Ochs, R.; Beckmann, D.; Weber, H. B.; Mayor, M.; L\"{o}hneysen, H. v. {\it Phys. Rev. Lett.} {\bf 2002}, {\it 88}, 176804.

\bibitem{Zhao} Zhao J.; Uosaki, K. {\it Nano Lett.} {\bf 2002}, {\it 2}, 137.

\bibitem{Wold} Wold, D. J.; Frisbie, C. D. {\it J. Am. Chem. Soc.} {\bf 2001}, {\it 123}, 5549; 
Wold, D. J.; Hagg, R.; Rampi, M. A.; Frisbie, C. D. {\it J. Phys. Chem. B} {\bf 2002}, {\it 106}, 2813.

\bibitem{Cui2} Cui, X. D.; Zarate, X.; Tomfohr, J.; Sankey, O. F.; Primak, A.; 
Moore, A. L.; Moore, T. A.; Gust, D.; Harris, G.; Lindsay, S. M. {\it Nanotechnology} {\bf 2002}, {\it 13}, 5.

\bibitem{Beebe} Beebe, J. M.; Engelkes, V. B.; Miller, L. L.; Frisbie, C. D. {\it J. Am. Chem. Soc.} {\bf 2002}, {\it 124}, 11268.

\bibitem{Holmlin} Holmlin, R.; Hagg, R.; Chabinyc, M. L.; Ismagilov, R. F.; Cohen, A. E.; Terfort, A.; Rampi, M. A.; Whitesides, G. M. 
{\it J. Am. Chem. Soc.} {\bf 2001}, {\it 123}, 5075.

\bibitem{Guo} Kaun, C.-C.; Guo, H. {\it Nano Lett.} {\bf 2003}, {\it 3}, 1521.

\bibitem{dicarlo} Pecchia, A.; Di Carlo, A.; Gagliardi, A.; Sanna, S.; Frauenheim, T.; Gutierrez, R. {\it Nano Lett.} {\bf 2004}, {\it 4}, 2109.

\bibitem{Cui} Cui, X. D.; Primak, A.; Zarate, X.; Tomfohr, J.; Sankey, O. F.; Moore, A. L.; Moore,  T. A.; Gust, D.; Harris, G.; Lindsay, S. M. 
{\it Science} {\bf 2001}, {\it 294}, 571.

\bibitem{Kushmerick} Kushmerick, J. G.; Lazorcik, J.; Patterson, C. H.; 
Shashidhar, R. {\it Nano Lett.} {\bf 2004}, {\it 4}, 639.

\bibitem{Wang1} Wang, W.; Lee, T.; Reed, M. A. {\it Phys. Rev. B} {\bf 2003}, {\it 68}, 035416.

\bibitem{Wang2} Wang, W.; Lee, T.; Kretzachmar, I.; Reed, M. A. {\it Nano Lett.} {\bf 2004}, {\it 4}, 643.

\bibitem{Kluth} Kluth, G. J.; Carraro, C.; Maboudian, R. {\it Phys. Rev. B} {\bf 1999}, {\it 59}, R10499.

\bibitem{DiVentra04} Yang, Z.; Chshiev, M.; Zwolak, M.; Chen, Y.-C.; Di Ventra, M. cond-mat/0409772.

\bibitem{Todorov2} M.J. Montgomery, J. Hoekstra, T.N Todorov, and A.P.
Sutton, J. Phys.: Cond. Mat. \textbf{15}, 731 (2003).

\bibitem{DiVentra03}  Chen, Y.-C.; Zwolak, M.; Di Ventra, M. {\it Nano Lett.} {\bf 2003}, {\it 3}, 1691.

\bibitem{smit} Smit, R. H. M.; Untiedt, C.; van Ruitenbeek, J. M. {\it Nanotechnology} {\bf 2004}, {\it 15}, S472.

\bibitem{Kato} Kato, H. S.; Noh, J.; Hara, M.; Kawai, M. {\it J. Phys. Chem. B} {\bf 2002}, {\it 106}, 9655.

\bibitem{cheninelastic} Chen, Y.-C.; Zwolak, M.; Di Ventra, M. {\it Nano Lett.}  
{\bf 2004}, {\it 4}, 1709.

\bibitem{diventra} Lang, N. D. {\it Phys. Rev. B} {\bf 1995}, {\it 52}, 5335; Di
Ventra M.; Lang, N. D. {\it Phys. Rev. B} {\bf 2002}, {\it 65}, 045402.

\bibitem{prec} Experiments have found an inverse decay length, $\beta$, of 
$\sim0.83$ to $\sim0.72$ $\mathring{A}^{-1}$, which is in good agreement with theoretical 
calculations.~\cite{Wang1,Guo} In our work, we use an average scaling factor of $\beta \simeq 0.78$ $\mathring{A}^{-1}$. 
The scaled current for octanethiolate at $0.1$ V is about $1.13$ nA, which is in good agreement with the value
calculated from previous DFT calculations.~\cite{Guo} Note, however, that a change in $\beta$ within the above range 
would not affect our conclusions. 

\bibitem{vibmodes} We have employed Hartree-Fock total energy calculations [see,
e.g., Boatz, J. A.; Gordon, M. S. {\it J. Phys. Chem.} {\bf 1989}, {\it 93}, 1819]
to evaluate the vibrational modes of the alkanethiolates with the sulfur atom
attached to a gold electrode. For these calculations, the gold
electrode is represented by a pad of five gold atoms with
infinite mass.
Four of the gold atoms are positioned at the corners of back-to-back (111) triangles, and the last gold atom is placed 
behind the triangles. The end atoms on the other contact form a CH$_3$ group. 
The end H atom is about 1.7\AA$\;$ away from the gold surface. The modes of this structure are not significantly 
influenced by the gold atoms of the second electrode surface. Note that it is well known that Hartree-Fock 
calculations overestimate the vibrational mode energies by about 10\% [see Scott, A. P.; Radom, L. {\it J. Phys. 
Chem.}, {\bf 1996}, {\it 100}, 16502]. Following this convention and for direct comparison with experiments, we multiply the energies of the modes by a factor of 0.9. 

\bibitem{Morikawa} Morikawa, Y.; Hayashi, T.; Liew, C. C.; Nozoye, H. {\it Surf. Sci.} {\bf 2002}, {\it 507}, 46.

\bibitem{Todorov1} Todorov, T. N. {\it Phil. Mag. B} {\bf 1998}, {\it 77}, 965; 
Montgomery, M. J.; Todorov, T. N.; Sutton, A. P. {\it J. Phys.: Cond. Mat.} {\bf 2002}, {\it 14}, 1.

\bibitem{geller} The stiffness is evaluated as $K=AY/d$, where $A\simeq 21.4$ $(a.u.)^{2}$ is the
effective cross section and $d$ is the effective length of the alkanethiol [see 
Patton, K. R.; Geller, M. R. {\it Phys. Rev. B} {\bf 2001}, {\it 64}, 155320]. The Young 
modulus, $Y\simeq 2.3\times 10^{12}$ $dyne/cm^{2}$, is calculated with total energy calculations 
and is found to be almost independent of the alkyl chain length. The spectral densities are estimated 
using the longitudinal and transverse sound velocities for gold, $v_{l}=3.2\cdot 10^{5}$ cm/sec and $v_{t}=1.2\cdot
10^{5}$ cm/sec, respectively. Note that a change in the thermal current by an order of magnitude would change the local temperature by just a factor of two. 

\bibitem{yang2003}  Yang,  Z.Q.; Di Ventra, M. {\it Phys. Rev. B} {\bf 2003}, {\it 67}, 161311. 

\bibitem{prec2} Note that the decreased strength of the average current-induced force per atom with increasing 
wire length has been shown for metallic junctions only [see, e.g., Refs.~\onlinecite{yang2003,DiVentra04}]. However, it is 
expected to be even more pronounced for insulating wires where the current decreases exponentially with 
length. 

\bibitem{troisi} Troisi, A.; Ratner, M. A.; Nitzan, A. {\it J. Chem. Phys.} {\bf 2003}, {\it 118}, 6072. 

\bibitem{Gronbeck} Gr\"onbeck, H.; Curioni, A.; Andreoni, W. {\it J. Am. Chem. Soc.} {\bf 2000}, {\it 122}, 3839.

\bibitem{Akinaga} Akinaga, Y.; Nakajima, T.; Hirao, K. {\it J. Chem. Phys.} {\bf 2001}, {\it 114}, 8555.

\bibitem{Yourdshahyan} Yourdshahyan, Y.; Zhang, H. K.; Rappe, A. M. {\it Phys. Rev. B} {\bf 2001}, {\it 63} R081405.

\bibitem{Vargas} Vargas, M. C.; Giannozzi, P.; Selloni, A.; Scoles, G. {\it J. Phys. Chem. B} {\bf 2001}, {\it 105} 9509.

\end{thebibliography}
\end{document}